\documentclass[12pt,aps,prd,showpacs,amsmath,amssymb]{revtex4}
\usepackage{graphicx}
\usepackage{epstopdf}
\usepackage{multirow}
\usepackage{subfigure}
\usepackage{appendix}
\input epsf
\begin{document}
\title{The magnetic moment of $Z_{c}(3900)$ as an axial-vector molecular state}
\author{Yong-Jiang Xu$^{1}$\footnote{xuyongjiang13@nudt.edu.cn}, Yong-Lu Liu$^1$, and Ming-Qiu Huang$^{1,2}$\footnote{corresponding author: mqhuang@nudt.edu.cn}}
\affiliation{$^1$Department of Physics, College of Liberal Arts and Sciences, National University of Defense Technology , Changsha, 410073, Hunan, China}
\affiliation{$^2$Synergetic Innovation Center for Quantum Effects and Applications, Hunan Normal University, Changsha,  410081, Hunan, China}
\date{}
\begin{abstract}
In this paper, we tentatively assign $Z_{c}(3900)$ to be an axialvector molecular state, and calculate its magnetic moment using the QCD sum rule method in external weak electromagnetic field. Starting with the two-point correlation function in external electromagnetic field and expanding it in power of the electromagnetic interaction Hamiltonian, we extract the mass and pole residue of $Z_{c}(3900)$ state from the leading term in the expansion and the magnetic moment from the linear response to the external electromagnetic field. The numerical values are $m_{Z_{c}}=3.97\pm0.12\mbox{GeV}$ in agreement with the experimental value $m^{exp}_{Z_{c}}=3899.0\pm3.6\pm4.9\mbox{MeV}$, $\lambda_{Z_{c}}=2.1\pm0.4\times10^{-2}\mbox{GeV}^{5}$ and $\mu_{Z_{c}}=0.19^{+0.04}_{-0.01}\mu_{N}$.
\end{abstract}
 \maketitle

\section{Introduction}\label{sec1}

$Z_{c}(3900)$, as a good candidate of exotic hadrons, was observed by BESIII collaboration in 2013 in the $\pi^{\pm}J/\psi$ invariant mass distribution of the process $e^{+}e^{-}\rightarrow\pi^{+}\pi^{-}J/\psi$ at a center-of-mass energy of $4.260\mbox{GeV}$ \cite{bes1}. Then the Belle and CLEO collaborations confirmed the existence of $Z_{c}(3900)$ \cite{belle,cleo}. In 2017, the BESIII collaboration determined the $J^{P}$ quantum number of $Z_{c}(3900)$ to be $J^{P}=1^{+}$ with a statistical significance larger than $7\sigma$ over other quantum numbers in a partial wave analysis of the process $e^{+}e^{-}\rightarrow\pi^{+}\pi^{-}J/\psi$ \cite{bes2}. Inspired by these experimental progress, there have been plentiful theoretical studies on $Z_{c}(3900)$'s properties through different approaches (see review article \cite{H.X.Chen} and references therein for details). However, the underlying structure of $Z_{c}(3900)$ is not understood completely and more endeavors are necessary in order to arrive at a better understanding for the properties of $Z_{c}(3900)$.

The electromagnetic multipole moments of hadron encode the spatial distributions of charge and magnetization in the hadron and provide important information about the quark configurations of the hadron and the underlying dynamics. So it is interesting to study the electromagnetic multipole moments of hadron.

The studies on the properties of hadrons inevitably involve the nonperturbative effects of quantum chromodynamics (QCD). The QCD sum rule method \cite{SVZ} is a nonperturbative analytic formalism firmly entrenched in QCD with minimal modeling and has been successfully applied in almost every aspect of strong interaction physics. In Ref.\cite{Balitsky,Ioffe1,Ioffe2}, the QCD sum rule method was extended to calculate the magnetic moments of the nucleon and hyperon in the external field method. In this method, a statics electromagnetic field is introduced which couples to the quarks and polarizes the QCD vacuum and magnetic moments can be extracted from the linear response to this field. Later, a more systematic studies was made for the magnetic moments of the octet baryons \cite{octet1,octet2,octet3,octet4}, the decuplet baryons \cite{decuplet1,decuplet2,decuplet3,decuplet4} and the $\rho$ meson \cite{rho}. In the case of the exotic X, Y, Z states, only the magnetic moment of $Z_{c}(3900)$ as an axialvector tetraquark state was calculated through this method \cite{wangzhigang}.

In this article, we study the magnetic moment of $Z_{c}(3900)$ as an axialvector molecular state with quantum number $J^{P}=1^{+}$ by the QCD sum rule method. The mass and pole residue, two of the input parameters needed to determine the magnetic moment, are calculated firstly including contributions of operators up to dimension 10. Then the magnetic moment is extracted from the linear term in $F_{\mu\nu}$ (external electromagnetic filed) of the correlation function.

The rest of the paper is arranged as follows. In Sec.\ref{sec2}, we derive the sum rules for the mass, pole residue and magnetic moment of $Z_{c}(3900)$ state. Sec.\ref{sec3} is devoted to the numerical analysis and a short summary is given in Sec.\ref{sec4}. In the Appendix \ref{appendix}, the spectral densities are shown.

\section{The derivation of the sum rules}\label{sec2}

The starting point of our calculation is the time-ordered correlation function in the QCD vacuum in the presence of a constant background electromagnetic field $F_{\mu\nu}$,
\begin{equation}\label{2-point correlator}
 \Pi_{\mu\nu}(p)=i\int dx^{4}e^{ipx}\langle0\mid\textsl{T}[J_{\mu}(x)J^{\dagger}_{\nu}(0)]\mid0\rangle_{F}
       =\Pi^{(0)}_{\mu\nu}(p)+\Pi^{(1)}_{\mu\nu\alpha\beta}(p)F^{\alpha\beta}+\cdots,
 \end{equation}
 where
 \begin{equation}\label{Zc interpolating current}
J_{\mu}(x)=\frac{1}{\sqrt{2}}\{[\bar{u}(x)i\gamma^{5}c(x)][\bar{c}(x)\gamma_{\mu}d(x)]+[\bar{u}(x)\gamma_{\mu}c(x)][\bar{c}(x)i\gamma^{5}d(x)]\}
\end{equation}
is the interpolating current of $Z_{c}(3900)$ as a molecular state with $J^{P}=1^{+}$ \cite{cuichunyu}. The $\Pi^{(0)}_{\mu\nu}(p)$ term is the correlation function without external electromagnetic field, and give rise to the mass and pole residue of $Z_{c}(3900)$. The magnetic moment will be extracted from the linear response term, $\Pi^{(1)}_{\mu\nu\alpha\beta}(p)F^{\alpha\beta}$.

The external electromagnetic field can interact directly with the quarks inside the hadron and also polarize the QCD vacuum. As a consequence, the vacuum condensates involved in the operator product expansion of the correlation function in the external electromagnetic field $F_{\mu\nu}$ are,
\begin{itemize}
  \item{dimension-2 operator},
  \begin{equation}
  F_{\mu\nu},
  \end{equation}
  \item{dimension-3 operator},
  \begin{equation}
  \langle0|\bar{q}\sigma_{\mu\nu}q|0\rangle_{F},
  \end{equation}
  \item{dimension-5 operators},
  \begin{equation}
  \langle0|\bar{q}q|0\rangle F_{\mu\nu}, \langle0|\bar{q}g_{s}G_{\mu\nu}q|0\rangle_{F}, \epsilon_{\mu\nu\alpha\beta}\langle0|\bar{q}g_{s}G^{\alpha\beta}q|0\rangle_{F},
  \end{equation}
  \item{dimension-6 operators},
  \begin{equation}
  \langle0|\bar{q}q|0\rangle\langle0|\bar{q}\sigma_{\mu\nu}q|0\rangle_{F}, \langle0|g^{2}_{s}GG|0\rangle F_{\mu\nu},\cdots,
  \end{equation}
  \item{dimension-7 operators},
  \begin{equation}
  \langle0|g^{2}_{s}GG|0\rangle\langle0|\bar{q}\sigma_{\mu\nu}q|0\rangle_{F}, \langle0| g_{s}\bar{q}\sigma\cdot Gq|0\rangle F_{\mu\nu}, \cdots,
  \end{equation}
  \item{dimension-8 operators},
  \begin{eqnarray}
  &&\langle0|\bar{q}q|0\rangle^{2}F_{\mu\nu}, \langle0| g_{s}\bar{q}\sigma\cdot Gq|0\rangle\langle0|\bar{q}\sigma_{\mu\nu}q|0\rangle_{F}, \langle0|\bar{q}q|0\rangle\langle0|\bar{q}g_{s}G_{\mu\nu}q|0\rangle_{F},\nonumber\\ &&\epsilon_{\mu\nu\alpha\beta}\langle0|\bar{q}q|0\rangle\langle0|\bar{q}g_{s}G^{\alpha\beta}q|0\rangle_{F},\cdots,
  \end{eqnarray}
\end{itemize}
The new vacuum condensates induced by the external electromagnetic field $F_{\mu\nu}$ can be described by introducing new parameters, $\chi$, $\kappa$ and $\xi$, called vacuum susceptibilities as follows,
\begin{eqnarray}
&&\langle0|\bar{q}\sigma_{\mu\nu}q|0\rangle_{F}=ee_{q}\chi\langle0|\bar{q}q|0\rangle F_{\mu\nu},\nonumber\\
&&\langle0|\bar{q}g_{s}G_{\mu\nu}q|0\rangle_{F}=ee_{q}\kappa\langle0|\bar{q}q|0\rangle F_{\mu\nu},\nonumber\\
&&\epsilon_{\mu\nu\alpha\beta}\langle0|\bar{q}g_{s}G^{\alpha\beta}q|0\rangle_{F}=iee_{q}\xi\langle0|\bar{q}q|0\rangle F_{\mu\nu}.
\end{eqnarray}

In order to express the two-point correlation function (\ref{2-point correlator}) physically, we expand it in powers of the electromagnetic interaction Hamiltonian $H_{int}=-ie\int d^{4}yj^{em}_{\alpha}(y)A^{\alpha}(y)$,
\begin{eqnarray}
\Pi_{\mu\nu}(p)=&&i\int dx^{4}e^{ipx}\langle0\mid\textsl{T}[J_{\mu}(x)J^{\dagger}_{\nu}(0)]\mid0\rangle\nonumber\\&&+i\int dx^{4}e^{ipx}\langle0\mid\textsl{T}\{J_{\mu}(x)[-ie\int d^{4}yj^{em}_{\alpha}(y)A^{\alpha}(y)]J^{\dagger}_{\nu}(0)\}\mid0\rangle+\cdots,
\end{eqnarray}
where $j^{em}_{\alpha}(y)$ is the electromagnetic current and $A^{\alpha}(y)$ is the electromagnetic four-vector.

Inserting complete sets of relevant states with the same quantum numbers as the current operator $J_{\mu}(x)$ and carrying out involved integrations, one has
\begin{eqnarray}\label{hadronic side}
\Pi^{had}_{\mu\nu}(p)=&&\frac{\lambda^{2}_{Z_{c}}}{m^{2}_{Z_{c}}-p^{2}}(-g_{\mu\nu}+\frac{p_{\mu}p_{\nu}}{p^{2}})
\nonumber\\&&-i\frac{\lambda^{2}_{Z_{c}}G_{2}(0)}{(p^{2}-m^{2}_{Z_{c}})^{2}}F_{\mu\nu}+i\frac{a}{m^{2}_{Z_{c}}-p^{2}}F_{\mu\nu}+\cdots,
\end{eqnarray}
where we make use of the following matrix elements
\begin{equation}
\langle0|J_{\mu}(0)|Z_{c}(p)\rangle=\lambda_{Z_{c}}\epsilon_{\mu}(p)
\end{equation}
with $\lambda_{Z_{c}}$ and $\epsilon_{\mu}(p)$ being the pole residue and polarization vector of $Z_{c}(3900)$, respectively,
\begin{eqnarray}
\langle Z_{c}(p)|j^{em}_{\alpha}(0)|Z_{c}(p^{\prime})\rangle=&&G_{1}(Q^{2})\epsilon^{*}(p)\cdot\epsilon(p^{\prime})
(p+p^{\prime})_{\alpha}+G_{2}(Q^{2})[\epsilon_{\alpha}(p^{\prime})\epsilon^{*}(p)\cdot q-\epsilon^{*}_{\alpha}(p)\epsilon(p^{\prime})\cdot q]\nonumber\\&&
-\frac{G_{3}(Q^{2})}{2m^{2}_{Z_{c}}}\epsilon^{*}(p)\cdot q\epsilon(p^{\prime})\cdot q(p+p^{\prime})_{\alpha}
\end{eqnarray}
with $q=p^{\prime}-p$ and $Q^{2}=-q^{2}$. The Lorentz-invariant functions $G_{1}(Q^{2})$, $G_{2}(Q^{2})$ and $G_{3}(Q^{2})$ are related to the charge, magnetic and quadrupole form-factors,
\begin{eqnarray}
&&G_{C}(Q^{2})=G_{1}(Q^{2})+\frac{2}{3}\eta G_{Q}(Q^{2}),\nonumber\\
&&G_{M}(Q^{2})=-G_{2}(Q^{2}),\nonumber\\
&&G_{Q}(Q^{2})=G_{1}(Q^{2})+G_{2}(Q^{2})+(1+\eta)G_{3}(Q^{2}),
\end{eqnarray}
respective, where $\eta=\frac{Q^{2}}{4m^{2}_{Z_{c}}}$. At zero momentum transfer, these form-factors are proportional to the usual static quantities of the charge $e$, magnetic moment $\mu_{Z_{c}}$ and quadrupole moment $Q_{1}$,
\begin{eqnarray}
&&eG_{C}(0)=e,\nonumber\\
&&eG_{M}(0)=2m_{Z}\mu_{Z_{c}},\nonumber\\
&&eG_{Q}(0)=m^{2}_{Z_{c}}Q_{1}.
\end{eqnarray}
The constant $a$ parameterizes the contributions from the pole-continuum transitions.

On the other hand, $\Pi_{\mu\nu}(p)$ can be calculated theoretically via OPE method at the quark-gluon level. To this end, one can insert the interpolating current $J_{\mu}(x)$ (\ref{Zc interpolating current}) into the correlation function (\ref{2-point correlator}), contract the relevant quark fields via Wick's theorem and obtain
\begin{eqnarray}
 \Pi^{OPE}_{\mu\nu}(p)=\frac{i}{2}\int d^{4}xe^{ipx}&&(Tr[(i\gamma_{5})S^{(u)}_{ca}(-x)(i\gamma_{5})S^{(c)}_{ac}(x)]Tr[\gamma_{\mu}S^{(d)}_{bd}(x)\gamma_{\nu}S^{(c)}_{db}(-x)]\nonumber\\
 &&+Tr[(i\gamma_{5})S^{(c)}_{db}(-x)\gamma_{\mu}S^{(d)}_{bd}(x)]Tr[(i\gamma_{5})S^{(c)}_{ac}(x)\gamma_{\nu}S^{(u)}_{ca}(-x)]\nonumber\\
 &&+Tr[(i\gamma_{5})S^{(u)}_{ca}(-x)\gamma_{\mu}S^{(c)}_{ac}(x)]Tr[(i\gamma_{5})S^{(d)}_{bd}(x)\gamma_{\nu}S^{(c)}_{db}(-x)]\nonumber\\
 &&+Tr[(i\gamma_{5})S^{(d)}_{bd}(x)(i\gamma_{5})S^{(c)}_{db}(-x)]Tr[\gamma_{\mu}S^{(c)}_{ac}(x)\gamma_{\nu}S^{(u)}_{ca}(-x)]),
\end{eqnarray}
where $S^{(c)}(x)=\langle 0|T[c(x)\bar{c}(0)]|0\rangle$ and $S^{(q)}(x)=\langle 0|T[q(x)\bar{q}(0)]|0\rangle, q=u, d$ are the full charm- and up (down)-quark propagators, whose expressions are given in the Appendix \ref{appendix1}, $Tr$ denotes the trace of the Dirac spinor indices, and $a$, $b$, $c$ and $d$ are color indices. Through dispersion relation, $\Pi^{OPE}_{\mu\nu}(p)$ can be written as
 \begin{equation}\label{QCD side}
 \Pi^{OPE}_{\mu\nu}(p)=\int^{\infty}_{4m^{2}_{c}}ds\frac{\rho^{(0)}(s)}{s-p^{2}}(-g_{\mu\nu}+\frac{p_{\mu}p_{\nu}}{p^{2}})
 +\int^{\infty}_{4m^{2}_{c}}ds\frac{\rho^{(1)}(s)}{s-p^{2}}(iF_{\mu\nu})+\mbox{other Lorentz structures},
 \end{equation}
 where $\rho^{i}(s)=\frac{1}{\pi}\mbox{Im}\Pi^{OPE}_{i}(s), i=0,1$ are the spectral densities. The spectral densities $\rho^{i}(s)$ are given in the Appendix \ref{appendix}.

 Finally, matching the phenomenological side (\ref{hadronic side}) and the QCD representation (\ref{QCD side}), we obtain
 \begin{equation}
 \frac{\lambda^{2}_{Z_{c}}}{m^{2}_{Z_{c}}-p^2}+\cdots=\int^{\infty}_{4m^{2}_{c}}ds\frac{\rho^{(0)}(s)}{s-p^2},
 \end{equation}
 for the Lorentz-structure $(-g_{\mu\nu}+\frac{p_{\mu}p_{\nu}}{p^{2}})$, and
 \begin{equation}
 \frac{\lambda^{2}_{Z_{c}}G_{M}(0)}{(m^{2}_{Z_{c}}-p^{2})^2}+\frac{a}{m^{2}_{Z_{c}}-p^2}+\cdots=\int^{\infty}_{4m^{2}_{c}}ds\frac{\rho^{(1)}(s)}{s-p^2},
 \end{equation}
 for the Lorentz-structure $iF_{\mu\nu}$.

 According to quark-hadron duality, the excited and continuum states' spectral density can be approximated by the QCD spectral density above some effective threshold $s^{0}_{Z_{c}}$, whose vale will be determined in Sec.\ref{sec3},
 \begin{eqnarray}
 &&\frac{\lambda^{2}_{Z_{c}}}{m^{2}_{Z_{c}}-p^2}+\int^{\infty}_{s^{0}_{Z_{c}}}ds\frac{\rho^{(0)}(s)}{s-p^2}=\int^{\infty}_{4m^{2}_{c}}ds\frac{\rho^{(0)}(s)}{s-p^2},\nonumber\\
 &&\frac{\lambda^{2}_{Z_{c}}G_{M}(0)}{(m^{2}_{Z_{c}}-p^{2})^2}+\frac{a}{m^{2}_{Z_{c}}-p^2}+\int^{\infty}_{s^{0}_{Z_{c}}}ds\frac{\rho^{(1)}(s)}{s-p^2}=\int^{\infty}_{4m^{2}_{c}}ds\frac{\rho^{(1)}(s)}{s-p^2}.
 \end{eqnarray}
 Subtracting the contributions of the excited and continuum states, one gets
 \begin{eqnarray}
 &&\frac{\lambda^{2}_{Z_{c}}}{m^{2}_{Z_{c}}-p^2}=\int^{s^{0}_{Z_{c}}}_{4m^{2}_{c}}ds\frac{\rho^{(0)}(s)}{s-p^2},\nonumber\\
 &&\frac{\lambda^{2}_{Z_{c}}G_{M}(0)}{(m^{2}_{Z_{c}}-p^{2})^2}+\frac{a}{m^{2}_{Z_{c}}-p^2}=\int^{s^{0}_{Z_{c}}}_{4m^{2}_{c}}ds\frac{\rho^{(1)}(s)}{s-p^2}.
 \end{eqnarray}

 In order to improve the convergence of the OPE series and suppress the contributions from the excited and continuum states, it is necessary to make a Borel transform. As a result, we have
 \begin{eqnarray}\label{2-point sum rule}
 &&\lambda^{2}_{Z_{c}}e^{-m^{2}_{Z_{c}}/M^{2}_{B}}=\int^{s^{0}_{Z_{c}}}_{4m^{2}_{c}}ds\rho^{(0)}(s)e^{-s/M^{2}_{B}},\nonumber\\
 &&\lambda^{2}_{Z_{c}}(\frac{G_{M}(0)}{M^{2}_{B}}+A)e^{-m^{2}_{Z_{c}}/M^{2}_{B}}=\int^{s^{0}_{Z_{c}}}_{4m^{2}_{c}}ds\rho^{(1)}(s)e^{-s/M^{2}_{B}}.
 \end{eqnarray}
 where $M^{2}_{B}$ is the Borel parameter and $A=\frac{a}{\lambda^{2}_{Z_{c}}}$. Taking derivative of the first equation in (\ref{2-point sum rule}) with respect to $-\frac{1}{M^{2}_{B}}$ and dividing it by the original expression, one has
 \begin{equation}\label{mass}
 m^{2}_{Z_{c}}=\frac{\frac{d}{d(-\frac{1}{M^{2}_{B}})}\int^{s^{0}_{Z_{c}}}_{4m^{2}_{c}}ds\rho^{(0)}(s)e^{-\frac{s}{M^{2}_{B}}}}{\int^{s^{0}_{Z_{c}}}_{4m^{2}_{c}}ds\rho^{(0)}(s)e^{-\frac{s}{M^{2}_{B}}}},
 \end{equation}
In the next section, (\ref{2-point sum rule}) and (\ref{mass}) will be analysed numerically to obtain the numerical values of the mass, the pole residue and the magnetic moment of the $Z_{c}(3900)$.

\section{Numerical analysis}\label{sec3}

The input parameters needed in numerical analysis are presented in Table \ref{input parameters}. For the vacuum susceptibilities $\chi$, $\kappa$ and $\xi$, we take the values $\chi=-(3.15\pm0.30)\mbox{GeV}^{-2}$, $\kappa=-0.2$ and $\xi=0.4$ determined in the detailed QCD sum rules analysis of the photon light-cone distribution amplitudes \cite{P.Ball}. Besides these parameters, we should determine the working intervals of the threshold parameter $s^{0}_{Z_{c}}$ and the Borel mass $M^{2}_{B}$ in which the mass, the pole residue and the magnetic moment vary weakly. The continuum threshold is related to the square of the first exited states having the same quantum number as the interpolating field, while the Borel parameter is determined by demanding that both the contributions of the higher states and continuum are sufficiently suppressed and the contributions coming from higher dimensional operators are small.
\begin{table}[htb]
\caption{Some input parameters needed in the calculations.}\label{input parameters}
\begin{tabular}{|c|c|}
  \hline
  Parameter      &   Value    \\
  \hline
  {$\langle\bar{q}q\rangle$}  &      $-(0.24\pm0.01)^{3}\mbox{GeV}^{3}$                     \\
  {$\langle g_{s}\bar{q}\sigma Gq\rangle$} & $(0.8\pm0.1)\langle\bar{q}q\rangle \mbox{GeV}^{2}$ \\
  {$\langle g^{2}_{s}GG\rangle$}    &     $0.88\pm0.25\mbox{GeV}^{4}$                \\
  {$m_{c}$}  &    $1.275^{+0.025}_{-0.035}\mbox{GeV}$\cite{M.Tanabashi}                   \\
  \hline
\end{tabular}
\end{table}

We define two quantities, the ratio of the pole contribution to the total contribution (RP) and the ratio of the highest dimensional term in the OPE series to the total OPE series (RH), as followings,
\begin{eqnarray}
&&RP_{i}\equiv\frac{\int^{s^{0}_{Z_{c}}}_{4m^{2}_{c}}ds\rho^{(i)}(s)e^{-\frac{s}{M^{2}_{B}}}}{\int^{\infty}_{4m^{2}_{c}}ds\rho^{(i)}(s)e^{-\frac{s}{M^{2}_{B}}}},
\nonumber\\&&RH_{i}\equiv\frac{\int^{s^{0}_{Z_{c}}}_{4m^{2}_{c}}ds\rho^{(d=n)}_{i}(s)e^{-\frac{s}{M^{2}_{B}}}}{\int^{s^{0}_{Z_{c}}}_{4m^{2}_{c}}ds\rho^{(i)}(s)e^{-\frac{s}{M^{2}_{B}}}},
\end{eqnarray}
where $i=0,1$ and $n=10(8)$ as $i=0(1)$, respectively.

\begin{figure}[htb]
\subfigure[]{
\includegraphics[width=7cm]{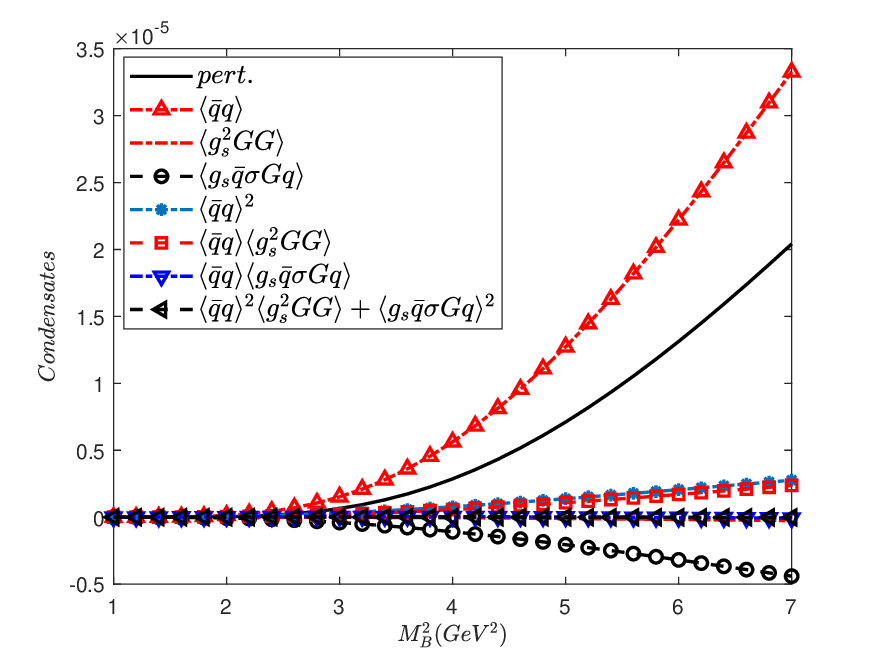}}
\subfigure[]{
\includegraphics[width=7cm]{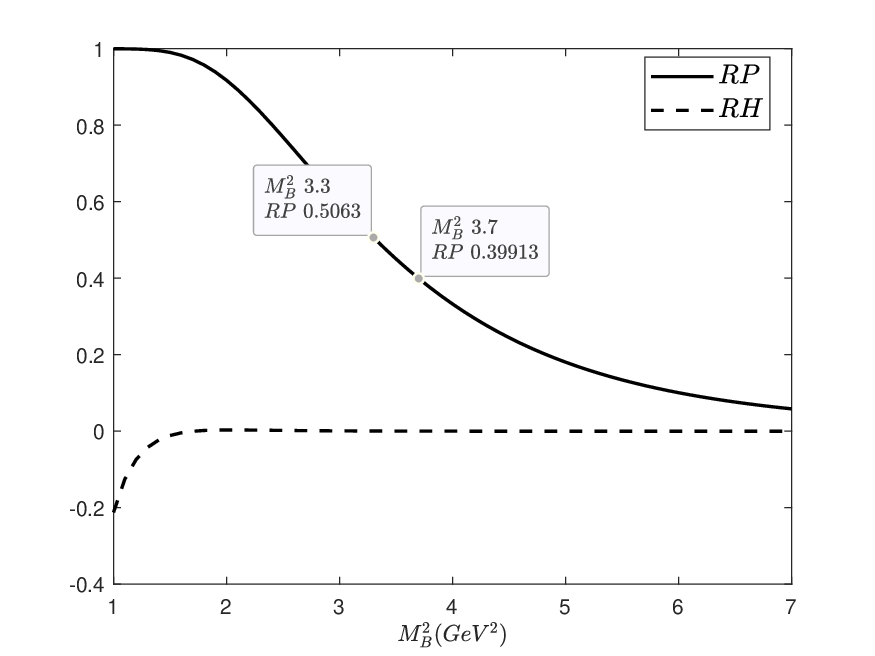}}
\caption{(a) denotes the various condensates as functions of $M^{2}_{B}$ with $\sqrt{s^{0}_{Z_{c}}}=4.6\mbox{GeV}$; (b) represents $RP_{0}$ and $RH_{0}$ varying with $M^{2}_{B}$ at $\sqrt{s^{0}_{Z_{c}}}=4.6\mbox{GeV}$.}\label{MB_range}
\end{figure}

In Fig.\ref{MB_range}(a), we compare the various terms in the OPE series as functions of $M^{2}_{B}$ with $\sqrt{s^{0}_{Z_{c}}}=4.6\mbox{GeV}$. From it one can see that except the quark condensate $\langle\bar{q}q\rangle$, other vacuum condensates are much smaller than the perturbative term. So the OPE series are under control. Fig.\ref{MB_range}(b) shows $RP_{0}$ and $RH_{0}$ varying with $M^{2}_{B}$ at $\sqrt{s^{0}_{Z_{c}}}=4.6\mbox{GeV}$. The figure shows that the requirement $RP_{0}\geq50\%$ ($RP_{0}\geq40\%$) gives $M^{2}_{B}\leq3.3\mbox{GeV}^{2}$ ($M^{2}_{B}\leq3.7\mbox{GeV}^{2}$) and $RH_{0}=5\%$ at $M^{2}_{B}=1.25\mbox{GeV}^{2}$.

\begin{figure}[htb]
\subfigure[]{
\includegraphics[width=7cm]{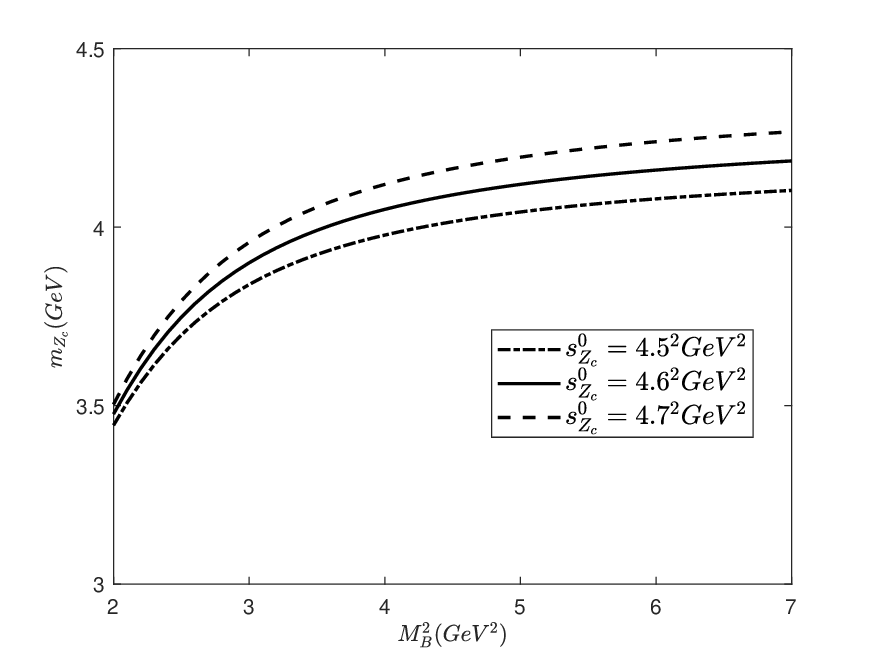}}
\subfigure[]{
\includegraphics[width=7cm]{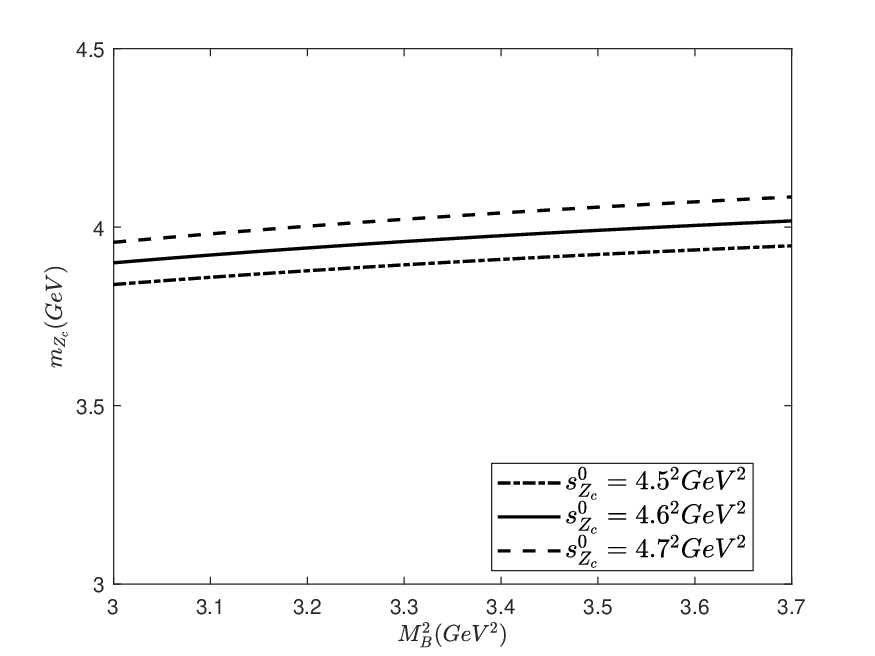}}
\caption{The dependence of the mass $m_{Z_{c}}$ on the Borel parameter $M^{2}_{B}$ with $\sqrt{s^{0}_{Z_{c}}}=4.5\mbox{GeV}$ (dot-dashed line), $\sqrt{s^{0}_{Z_{c}}}=4.6\mbox{GeV}$ (real line) and $\sqrt{s^{0}_{Z_{c}}}=4.6\mbox{GeV}$ (dashed line).}\label{massfig}
\end{figure}

 From Fig.\ref{massfig}(a), we know that the sum rule for the mass $m_{Z_{c}}$ depends strongly on the Borel parameter $M^{2}_{B}$ as $M^{2}_{B}\leq3\mbox{GeV}^{2}$. Along with the criterions of pole dominance, this fact confines $M^{2}_{B}$ from $3\mbox{GeV}^{2}$ to $3.7\mbox{GeV}^{2}$. In the analysis, we take $RP_{0}\geq40\%$ so that we can obtain a larger interval of the Borel parameter. Within the interval of $M^{2}_{B}$ determined above, the mass varies weakly with $M^{2}_{B}$ as depicted in Fig.\ref{massfig}(b). Fig.\ref{massfig}(b) also shows the weak dependence of the mass on the threshold parameter $s^{0}_{Z_{c}}$ as $4.5^{2}\mbox{GeV}^{2}\leq s^{0}_{Z_{c}}\leq4.7^{2}\mbox{GeV}^{2}$. As a result, we can reliably read the value of the mass, $m_{Z_{c}}=3.97\pm0.12\mbox{GeV}$, in agreement with the experimental value $m^{exp}_{Z_{c}}=3899.0\pm3.6\pm4.9\mbox{MeV}$.

\begin{figure}[htb]
\includegraphics[width=10cm]{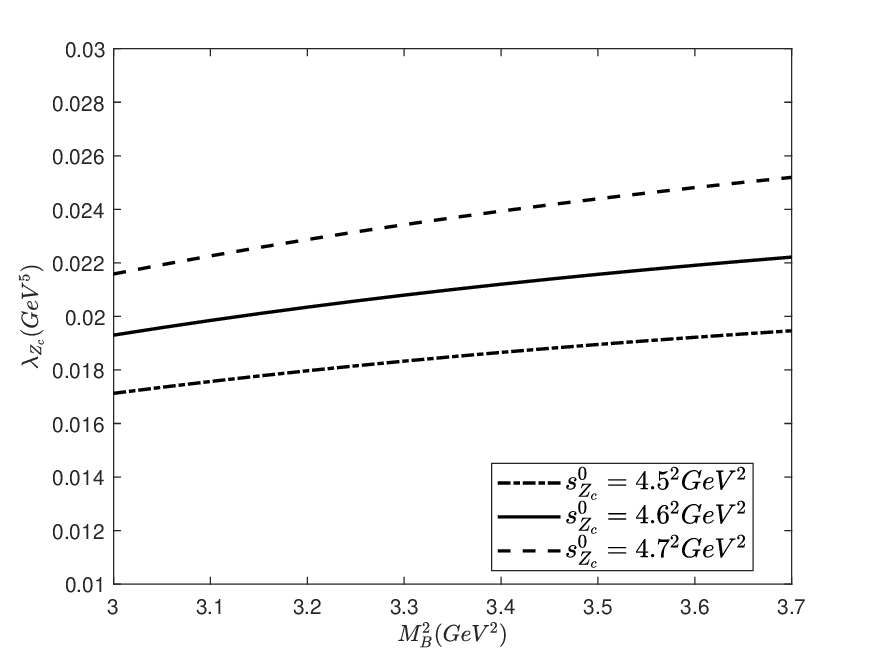}
\caption{The figure shows the dependence of the pole residue $\lambda_{Z_{c}}$ on the Borel parameter $M^{2}_{B}$ in the determined interval at three different values of $s^{0}_{Z_{c}}$.}\label{coupling}
\end{figure}

In Fig.\ref{coupling}, we show the variation of the pole residue with the Borel parameter $M^{2}_{B}$ in the determined interval at three different values of $s^{0}_{Z_{c}}$. It is obvious that the pole residue depends weakly on $M^{2}_{B}$ and $s^{0}_{Z_{c}}$ and $\lambda_{Z_{c}}=2.1\pm0.4\times10^{-2}\mbox{GeV}^{5}$.

\begin{figure}[htb]
\subfigure[]{
\includegraphics[width=7cm]{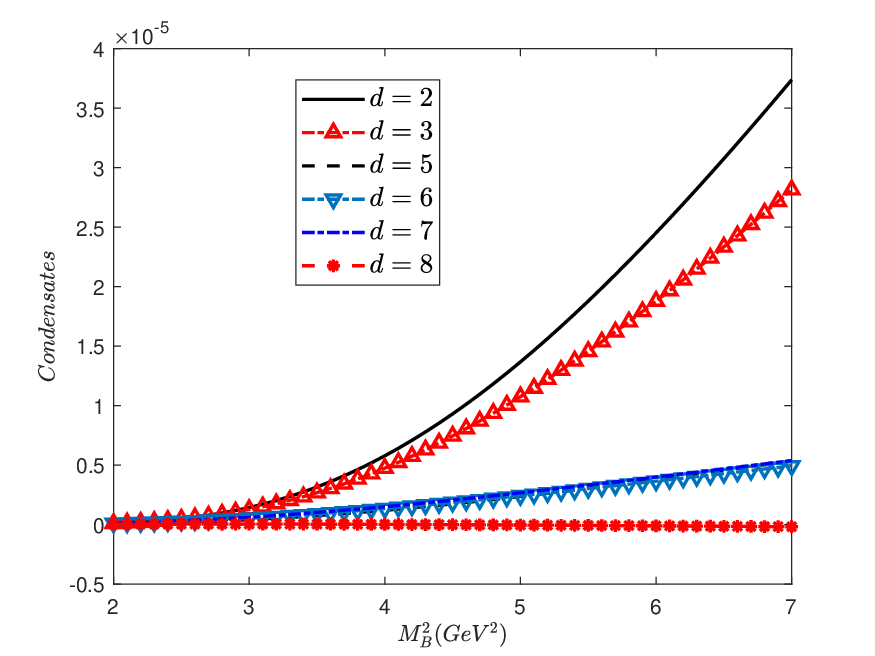}}
\subfigure[]{
\includegraphics[width=7cm]{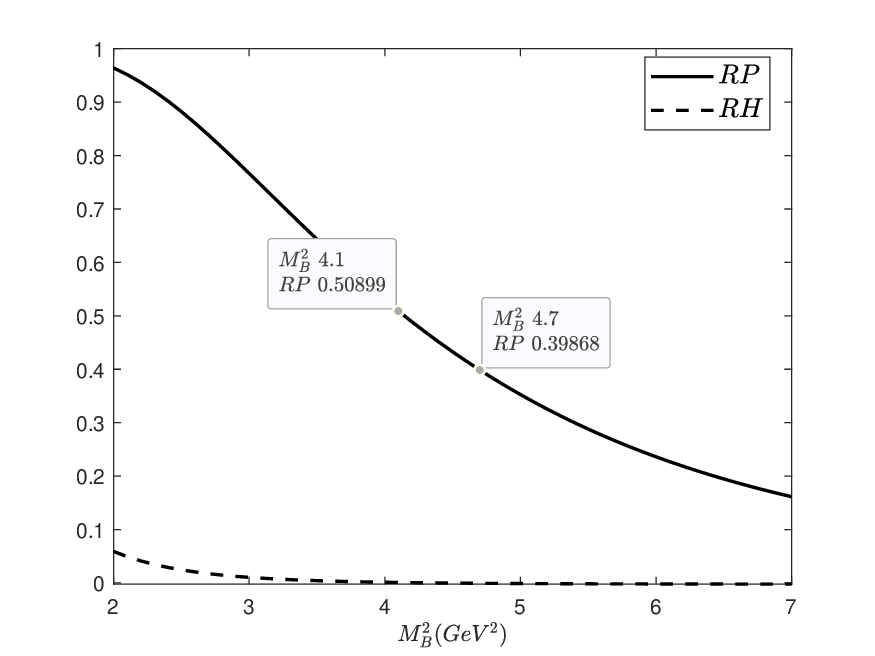}}
\subfigure[]{
\includegraphics[width=7cm]{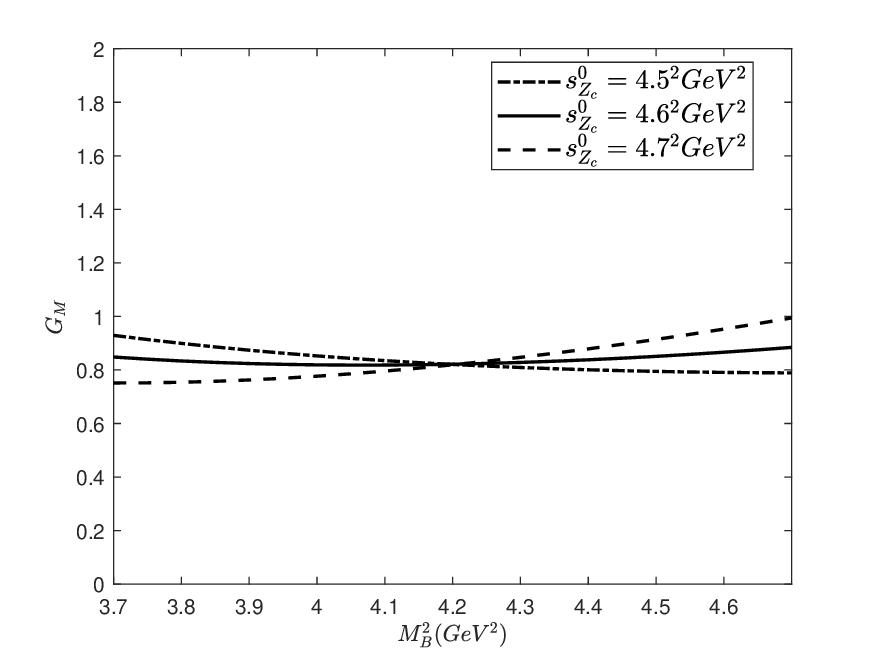}}
\caption{(a) shows the various condensates as functions of $M^{2}_{B}$ with $\sqrt{s^{0}_{Z_{c}}}=4.6\mbox{GeV}$; (b) presents $RP_{1}$ and $RH_{1}$ varying with $M^{2}_{B}$ at $\sqrt{s^{0}_{Z_{c}}}=4.6\mbox{GeV}$; (c) depicts the dependence of $G_{M}$ on $M^{2}_{B}$ in the determined interval at three different values of $s^{0}_{Z_{c}}$.}\label{magneticmoment}
\end{figure}

The same procedure can be done for the sum rule of the magnetic moment. The results are shown in Fig.\ref{magneticmoment}, from which the value of $G_{M}(0)$ can be read as $G_{M}(0)=0.82^{+0.17}_{-0.07}$. Finally, we obtain
\begin{equation}
\mu_{Z_{c}}=G_{M}(0)\frac{e}{2m_{Z_{c}}}=0.82^{+0.17}_{-0.07}\frac{e}{2m_{Z_{c}}}=0.19^{+0.04}_{-0.01}\mu_{N},
\end{equation}
where $\mu_{N}$ is the nucleon magneton.

In Ref.\cite{wangzhigang}, the author gave $\mu_{Z_{c}}=0.47^{+0.27}_{-0.22}\mu_{N}$ assuming $Z_{c}(3900)$ as an axialvector tetraquark state by the same method used in this article. In Ref.\cite{U.Ozdem}, $\mu_{Z_{c}}=0.67\pm0.32\mu_{N}$ was predicted using light-cone sum rule under the axialvector tetraquark assumption. In Table \ref{comparison}, we summarize the values of the magnetic moment of $Z_{c}(3900)$ under different assumptions about the quark configuration and with different methods. It is obvious that the magnetic moment of $Z_{c}(3900)$ has different values if $Z_{c}(3900)$ has different quark configurations. The theoretical predictions can be confronted to the experimental data in the future and give important information about the inner structure of $Z_{c}(3900)$ state.
\begin{table}[htb]
\caption{The magnetic moment of $Z_{c}(3900)$($\mu_{N}$ is the nucleon magneton).}\label{comparison}
\begin{tabular}{|c|c|c|}
  \hline
  Quark Configuration     &   Method     &  Value  \\
  \hline
  Axialvector Tetraquark  &  Light-Cone Sum Rule  &  $0.67\pm0.32\mu_{N}$\cite{U.Ozdem}  \\
  \hline
  Axialvector Tetraquark  &  QCD Sum Rule  &  $0.47^{+0.27}_{-0.22}\mu_{N}$\cite{wangzhigang}  \\
  \hline
  Axialvector Molecule    &  QCD Sum Rule  &   $0.19^{+0.04}_{-0.01}\mu_{N}$(this work) \\
  \hline
\end{tabular}
\end{table}

\section{Conclusion}\label{sec4}

In this paper, we tentatively assign $Z_{c}(3900)$ to be an axialvector molecular state, calculate its magnetic moment using the QCD sum rule method in the external weak electromagnetic field. Starting with the two-point correlation function in the external electromagnetic field and expanding it in power of the electromagnetic interaction Hamiltonian, we extract the mass and pole residue of $Z_{c}(3900)$ state from the leading term in the expansion and the magnetic moment from the linear response to the external electromagnetic field. The numerical values are $m_{Z_{c}}=3.97\pm0.12\mbox{GeV}$ in agreement with the experimental value $m^{exp}_{Z_{c}}=3899.0\pm3.6\pm4.9\mbox{MeV}$, $\lambda_{Z_{c}}=2.1\pm0.4\times10^{-2}\mbox{GeV}^{5}$ and $\mu_{Z_{c}}=0.19^{+0.04}_{-0.01}\mu_{N}$ with $\mu_{N}$ the nucleon magneton. The prediction can be confronted to the experimental data in the future and give important information about the inner structure of $Z_{c}(3900)$ state.

\acknowledgments  This work was supported by the National
Natural Science Foundation of China under Contract No.11675263.

\begin{appendix}
\section{The quark propagators}\label{appendix1}
The full quark propagators are
\begin{eqnarray}
 S^{q}_{ij}(x)=&&\frac{i \not\!{x}}{2\pi^{2}x^4}\delta_{ij}-\frac{m_{q}}{4\pi^2x^2}\delta_{ij}-\frac{\langle\bar{q}q\rangle}{12}\delta_{ij}
 +i\frac{\langle\bar{q}q\rangle}{48}m_{q}\not\!{x}\delta_{ij}-\frac{x^2}{192}\langle g_{s}\bar{q}\sigma Gq\rangle \delta_{ij}\nonumber\\
 &&+i\frac{x^2\not\!{x}}{1152}m_{q}\langle g_{s}\bar{q}\sigma Gq\rangle \delta_{ij}-i\frac{g_{s}t^{a}_{ij}G^{a}_{\mu\nu}}{32\pi^2x^2}(\not\!{x}\sigma^{\mu\nu}+\sigma^{\mu\nu}\not\!{x})\nonumber\\
 &&+i\frac{\delta_{ij}e_{q}F_{\mu\nu}}{32\pi^2x^2}(\not\!{x}\sigma^{\mu\nu}+\sigma^{\mu\nu}\not\!{x})
 -\frac{\delta_{ij}e_{q}\chi\langle\bar{q}q\rangle\sigma^{\mu\nu}F_{\mu\nu}}{24}\nonumber\\
 &&+\frac{\delta_{ij}e_{q}\langle\bar{q}q\rangle F_{\mu\nu}}{288}(\sigma^{\mu\nu}-2\sigma^{\alpha\mu}x_{\alpha}x^{\nu})\nonumber\\
 &&+\frac{\delta_{ij}e_{q}\langle\bar{q}q\rangle F_{\mu\nu}}{576}[(\kappa+\xi)\sigma^{\mu\nu}x^{2}-(2\kappa-\xi)\sigma^{\alpha\mu}x_{\alpha}x^{\nu}]+\cdots
 \end{eqnarray}
 for light quarks, and
 \begin{eqnarray}
 S^{Q}_{ij}(x)=i\int\frac{d^{4}k}{(2\pi)^4}e^{-ikx}&&[\frac{\not\!{k}+m_{Q}}{k^2-m^{2}_{Q}}\delta_{ij}
 -\frac{g_{s}t^{a}_{ij}G^{a}_{\mu\nu}}{4}\frac{\sigma^{\mu\nu}(\not\!{k}+m_{Q})+(\not\!{k}+m_{Q})\sigma^{\mu\nu}}
 {(k^2-m^{2}_{Q})^{2}}\nonumber\\
 &&+\frac{\langle g^{2}_{s}GG\rangle}{12}\delta_{ij}m_{Q}\frac{k^2+m_{Q}\not\!{k}}{(k^2-m^{2}_{Q})^{4}}\nonumber\\
 &&+\frac{\delta_{ij}e_{Q}F_{\mu\nu}}{4}\frac{\sigma^{\mu\nu}(\not\!{k}+m_{Q})+(\not\!{k}+m_{Q})\sigma^{\mu\nu}}
 {(k^2-m^{2}_{Q})^{2}}+\cdots]
 \end{eqnarray}
 for heavy quarks. In these expressions $t^{a}=\frac{\lambda^{a}}{2}$ and $\lambda^{a}$ are the Gell-Mann matrix, $g_{s}$ is the strong interaction coupling constant, and $i, j$ are color indices, $e_{Q(q)}$ is the charge of the heavy (light) quark and $F_{\mu\nu}$ is the external electromagnetic field.

\section{The spectral densities}\label{appendix}
On the QCD side, we carry out the OPE up to dimension-10 and dimension-8 for the spectral densities $\rho^{(0)}(s)$ and $\rho^{(1)}(s)$ respectively. The explicit expressions of the spectral densities are given below.

\begin{equation}
\rho^{(0)}(s)=\rho^{(d=0)}_{0}+\rho^{(d=3)}_{0}(s)+\rho^{(d=4)}_{0}(s)+\rho^{(d=5)}_{0}(s)+\rho^{(d=6)}_{0}(s)
+\rho^{(d=7)}_{0}(s)+\rho^{(d=8)}_{0}(s)+\rho^{(d=10)}_{0}(s),
\end{equation}
with
\begin{equation}
\rho^{(d=0)}_{0}(s)=\frac{3}{4096\pi^{6}}\int^{a_{max}}_{a_{min}}da\int^{1-a}_{b_{min}}db\frac{1}{a^{3}b^{3}}(1-a-b)(1+a+b)(m^{2}_{c}(a+b)-abs)^{4},
\end{equation}
\begin{equation}
\rho^{(d=3)}_{0}(s)=-\frac{3m_{c}\langle0|\bar{q}q|0\rangle}{256\pi^{4}}\int^{a_{max}}_{a_{min}}da\int^{1-a}_{b_{min}}db\frac{1}{a^{2}b^{2}}(a+b)(1+a+b)(m^{2}_{c}(a+b)-abs)^{2},
\end{equation}
\begin{eqnarray}
\rho^{(d=4)}_{0}(s)=&&\frac{m^{2}_{c}\langle0| g^{2}_{s}GG|0\rangle}{4096\pi^{6}}\int^{a_{max}}_{a_{min}}da\int^{1-a}_{b_{min}}db\frac{1}{a^{3}b^{3}}(a^{3}+b^{3})(1-a-b)(1+a+b)(m^{2}_{c}(a+b)-abs)\nonumber\\
&&+\frac{\langle0| g^{2}_{s}GG|0\rangle}{4096\pi^{6}}\int^{a_{max}}_{a_{min}}da\int^{1-a}_{b_{min}}db\frac{1}{a^{2}b^{2}}(a+b)(2a+2b-1)(m^{2}_{c}(a+b)-abs)^{2},
\end{eqnarray}
\begin{eqnarray}
\rho^{(d=5)}_{0}(s)=&&\frac{3m_{c}\langle0| g_{s}\bar{q}\sigma\cdot Gq|0\rangle}{256\pi^{4}}\int^{a_{max}}_{a_{min}}da\int^{1-a}_{b_{min}}db\frac{1}{a^{2}b^{2}}(a^{2}+b^{2})(a+b)(m^{2}_{c}(a+b)-abs)\nonumber\\
&&+\frac{3m_{c}\langle0| g_{s}\bar{q}\sigma\cdot Gq|0\rangle}{512\pi^{4}}\int^{a_{max}}_{a_{min}}da\int^{1-a}_{b_{min}}db\frac{1}{ab}(a+b)(m^{2}_{c}(a+b)-abs)\nonumber\\
&&-\frac{3m_{c}\langle0| g_{s}\bar{q}\sigma\cdot Gq|0\rangle}{256\pi^{4}}\int^{a_{max}}_{a_{min}}da\frac{1}{a(1-a)}(m^{2}_{c}-a(1-a)s),
\end{eqnarray}
\begin{equation}
\rho^{(d=6)}_{0}(s)=\frac{m^{2}_{c}\langle0|\bar{q}q0|\rangle^{2}}{16\pi^{2}}\frac{\sqrt{s(s-4m^{2}_{c})}}{s},
\end{equation}
\begin{eqnarray}
\rho^{(d=7)}_{0}(s)=&&-\frac{m_{c}\langle0|\bar{q}q0|\rangle\langle0| g^{2}_{s}GG|0\rangle}{512\pi^{4}}\int^{a_{max}}_{a_{min}}da\int^{1-a}_{b_{min}}db\frac{1}{a^2}b(1+a+b)\nonumber\\
&&-\frac{m_{c}\langle0|\bar{q}q|0\rangle\langle0| g^{2}_{s}GG|0\rangle}{1536\pi^{4}}\int^{a_{max}}_{a_{min}}da(1+a+b_{min})\frac{as+m^{2}_{c}}{m^{2}_{c}},
\end{eqnarray}
\begin{equation}
\rho^{(d=8)}_{0}(s)=\frac{\langle0|\bar{q}q|0\rangle\langle0| g_{s}\bar{q}\sigma\cdot Gq|0\rangle}{32\pi^2}\frac{m^{2}_{c}}{\sqrt{s(s-4m^{2}_{c})}}(2\frac{m^{2}_{c}}{M^{2}_{B}}+2\frac{m^{2}_{c}}{s}-1),
\end{equation}
\begin{eqnarray}
\rho^{(d=10)}_{0}(s)=&&\frac{\langle0|\bar{q}q|0\rangle^{2}\langle0| g^{2}_{s}GG|0\rangle}{576\pi^{2}}\frac{1}{\sqrt{s(s-4m^{2}_{c})}}(\frac{s(s-3m^{2}_{c})}{M^{4}_{B}}-3\frac{s-2m^{2}_{c}}{M^{2}_{B}})\nonumber\\
&&+\frac{\langle0| g_{s}\bar{q}\sigma\cdot Gq|0\rangle^{2}}{128\pi^{2}}\frac{m^{2}_{c}s}{M^{4}_{B}\sqrt{s(s-4m^{2}_{c})}}(1-\frac{m^{2}_{c}}{M^{2}_{B}}).
\end{eqnarray}

\begin{equation}
\rho^{(1)}(s)=\rho^{(d=2)}_{1}+\rho^{(d=3)}_{1}(s)+\rho^{(d=5)}_{1}(s)+\rho^{(d=6)}_{1}(s)
+\rho^{(d=7)}_{1}(s)+\rho^{(d=8)}_{1}(s),
\end{equation}
with
\begin{equation}
\rho^{(d=2)}_{1}(s)=-\frac{3}{1024\pi^{6}}\int^{a_{max}}_{a_{min}}da\int^{1-a}_{b_{min}}db\frac{1}{a^{2}b^{2}}(a+b)(m^{2}_{c}(a+b)-abs)^{3},
\end{equation}
\begin{equation}
\rho^{(d=3)}_{1}(s)=\frac{3m_{c}\chi\langle0|\bar{q}q|0\rangle}{256\pi^{4}}\int^{a_{max}}_{a_{min}}da\int^{1-a}_{b_{min}}db\frac{1}{ab^{2}}(m^{2}_{c}(a+b)-abs)^{2},
\end{equation}
\begin{eqnarray}
\rho^{(d=5)}_{1}(s)=&&-\frac{m_{c}\langle0|\bar{q}q0|\rangle}{32\pi^{4}}\int^{a_{max}}_{a_{min}}da\int^{1-a}_{b_{min}}db\frac{1}{b}(m^{2}_{c}(a+b)-abs)\nonumber\\
&&-\frac{m_{c}(2\kappa-\xi)\langle0|\bar{q}q|0\rangle}{512\pi^{4}}\int^{a_{max}}_{a_{min}}da\int^{1-a}_{b_{min}}db\frac{1}{b}(m^{2}_{c}(a+b)-abs)\nonumber\\
&&+\frac{3m_{c}(2\kappa+\xi)\langle0|\bar{q}q|0\rangle}{256\pi^{4}}\int^{a_{max}}_{a_{min}}da\int^{1-a}_{b_{min}}db\frac{1}{b^{2}}(m^{2}_{c}(a+b)-abs)\nonumber\\
&&+\frac{m_{c}\langle0|\bar{q}q|0\rangle}{64\pi^{4}}\int^{a_{max}}_{a_{min}}da\frac{1}{a}(m^{2}_{c}-a(1-a)s)\nonumber\\
&&-\frac{m_{c}(\kappa+\xi)\langle0|\bar{q}q|0\rangle}{256\pi^{4}}\int^{a_{max}}_{a_{min}}da\frac{1}{a}(m^{2}_{c}-a(1-a)s)
\end{eqnarray}
\begin{eqnarray}
\rho^{(d=6)}_{1}(s)=&&-\frac{m^{2}_{c}\langle0| g^{2}_{s}GG|0\rangle}{2048\pi^{6}}\int^{a_{max}}_{a_{min}}da\int^{1-a}_{b_{min}}db\frac{1}{b^{2}}a(a+b)\nonumber\\
&&+\frac{3\langle0| g^{2}_{s}GG|0\rangle}{4096\pi^{6}}\int^{a_{max}}_{a_{min}}da\int^{1-a}_{b_{min}}db\frac{1}{b}(m^{2}_{c}(a+b)-abs)\nonumber\\
&&-\frac{3\langle0| g^{2}_{s}GG|0\rangle}{4096\pi^{6}}\int^{a_{max}}_{a_{min}}da\frac{1}{a}(m^{2}_{c}-a(1-a)s)\nonumber\\
&&-\frac{m^{2}_{c}\chi\langle0|\bar{q}q|0\rangle^{2}}{32\pi^{2}}\frac{\sqrt{s(s-4m^{2}_{c})}}{s},
\end{eqnarray}
\begin{eqnarray}
\rho^{(d=7)}_{1}(s)=&&\frac{m_{c}\chi\langle0|\bar{q}q|0\rangle\langle0| g^{2}_{s}GG|0\rangle}{1024\pi^{4}}\int^{a_{max}}_{a_{min}}da\int^{1-a}_{b_{min}}db\frac{a}{b^{2}}\nonumber\\
&&+\frac{\chi\langle0|\bar{q}q|0\rangle\langle0| g^{2}_{s}GG|0\rangle}{3072\pi^{4}}\int^{a_{max}}_{a_{min}}da\frac{as+3m^{2}_{c}}{m_{c}}\nonumber\\
&&+\frac{3m_{c}\langle0|g_{s}\bar{q}\sigma\cdot Gq0|\rangle}{256\pi^{4}}\int^{a_{max}}_{a_{min}}da\int^{1-a}_{b_{min}}db\frac{a}{b}\nonumber\\
&&-\frac{3m_{c}\langle0|g_{s}\bar{q}\sigma\cdot Gq0|\rangle}{256\pi^{4}}\int^{a_{max}}_{a_{min}}da\frac{1-a}{a}\nonumber\\
&&-\frac{3m^{3}_{c}\langle0|g_{s}\bar{q}\sigma\cdot Gq0|\rangle}{512\pi^{4}}\frac{1}{\sqrt{s(s-4m^{2}_{c})}}\nonumber\\
&&+\frac{3m_{c}\langle0|g_{s}\bar{q}\sigma\cdot Gq0|\rangle}{1024\pi^{4}}\frac{\sqrt{s(s-4m^{2}_{c})}}{s},
\end{eqnarray}
\begin{eqnarray}
\rho^{(d=8)}_{1}(s)=&&-\frac{m^{2}_{c}\chi\langle0|\bar{q}q|0\rangle\langle0|g_{s}\bar{q}\sigma\cdot Gq0|\rangle}{64\pi^{2}}\frac{1}{\sqrt{s(s-4m^{2}_{c})}}(\frac{m^{2}_{c}}{M^{2}_{B}}+\frac{m^{2}_{c}}{s}-1)\nonumber\\
&&+\frac{m^{4}_{c}\langle0|\bar{q}q|0\rangle^{2}}{48\pi^{2}}\frac{1}{\sqrt{s(s-4m^{2}_{c})}}(\frac{1}{M^{2}_{B}}+\frac{2}{s})\nonumber\\
&&+\frac{m^{4}_{c}(2\kappa-\xi)\langle0|\bar{q}q|0\rangle^{2}}{192\pi^{2}}\frac{1}{s\sqrt{s(s-4m^{2}_{c})}}\nonumber\\
&&+\frac{m^{4}_{c}(\kappa+\xi)\langle0|\bar{q}q|0\rangle^{2}}{96\pi^{2}}\frac{1}{\sqrt{s(s-4m^{2}_{c})}}(\frac{1}{M^{2}_{B}}+\frac{1}{s})\nonumber\\
&&-\frac{m^{2}_{c}(2\kappa+\xi)\langle0|\bar{q}q|0\rangle^{2}}{64\pi^{2}}\frac{1}{\sqrt{s(s-4m^{2}_{c})}}.
\end{eqnarray}
In the above equations, $a_{max}=\frac{1+\sqrt{1-\frac{4m^{2}_{c}}{s}}}{2}$, $a_{min}=\frac{1-\sqrt{1-\frac{4m^{2}_{c}}{s}}}{2}$ and $b_{min}=\frac{am^{2}_{c}}{as-m^{2}_{c}}$.
\end{appendix}



\begin{thebibliography}{99}
\bibitem{bes1}M. Ablikim et al, Phys. Rev. Lett. 110 (2013) 252001.
\bibitem{belle}Z. Q. Liu et al, Phys. Rev. Lett. 110 (2013) 252002.
\bibitem{cleo}T. Xiao, S. Dobbs, A. Tomaradze and K. K. Seth, Phys. Lett. B727 (2013) 366.
\bibitem{bes2}M. Ablikim et al, Phys. Rev. Lett. 119 (2017) 072001.
\bibitem{H.X.Chen}Y. R. Liu, H. X. Chen, W. Chen, X. Liu, and S. L. Zhu, Prog. Part. Nucl. Phys. 107 (2019) 237; H. X. Chen, W. Chen, X. Liu, Y. R. Liu and S. L. Zhu, Rept. Prog. Phys. 80 (2017) no. 7, 076201; H. X. Chen, W. Chen, X. Liu and S. L. Zhu, Phys. Rept. 639 (2016) 1.
\bibitem{SVZ}M. A. Shifman, A. I. Vainshtein and V. I. Zakharov, Nucl. Phys. B147 (1979) 385; Nucl. Phys. B147 (1979) 448.
\bibitem{Balitsky}I. I. Balitsky and A. V. Yung, Phys. Lett. B129 (1983) 328.
\bibitem{Ioffe1}B. L. Ioffe and A. V. Smilga, Nucl. Phys. B232 (1984) 109.
\bibitem{Ioffe2}B. L. Ioffe and A. V. Smilga, Phys. Lett. B133 (1983) 436.
\bibitem{octet1}C. B. Chiu, J. Pasupathy and S. L. Wilson, Phys. Rev. D33 (1986) 1961.
\bibitem{octet2}J. Pasupathy, J. P. Singh, S. L. Wilson and C. B. Chiu, Phys. Rev. D36 (1986) 1442.
\bibitem{octet3}S. L. Wilson, J. Pasupathy and C. B. Chiu, Phys. Rev. D36 (1987) 1451.
\bibitem{octet4}S. Zhu, W. Hwang and Z. Yang, Phys. Rev. D57 (1998) 1527.
\bibitem{decuplet1}F. X. Lee, Phys. Rev. D57 (1998) 1801.
\bibitem{decuplet2}F. X. Lee, Phys. Lett. B419 (1998) 14.
\bibitem{decuplet3}J. Dey, M. Dey and A. Iqubal, Phys. Lett. B477 (2000) 125.
\bibitem{decuplet4}M. Sinha, A. Iqubal, M. Dey and J. Dey, Phys. Lett. B610 (2005) 283.
\bibitem{rho}A. Samsonov, Phys. Atom. Nucl. 68 (2005) 114.
\bibitem{wangzhigang}Z. G. Wang, Eur. Phys. J. C78 (2018)4 297.
\bibitem{cuichunyu}C. Y. Cui, X. H. Liao, Y. L. Liu and M. Q. Huang, J. Phys. G41 (2014) 075003.
\bibitem{M.Tanabashi}M. Tanabashi et al[Particle Data Group], Phys. Rev. D98 (2018) 030001.
\bibitem{P.Ball}P. Ball, V. M. Braun and N. Kivel, Nucl. Phys. B649 (2003) 263.
\bibitem{U.Ozdem}U. Ozdem and K. Azizi, Phys. Rev. D96 (2017) 074030.
\end{thebibliography}
\end{document}